\documentclass{birkjour}

 \usepackage{graphicx}   
 \usepackage{multicol}   
\usepackage{cite}
\usepackage{xcolor}
\usepackage{soul}
\makeindex             
\theoremstyle{plain}

\theoremstyle{definition}

\theoremstyle{remark}

\theoremstyle{definition}

\theoremstyle{definition}

\theoremstyle{remark}

\theoremstyle{definition}

\theoremstyle{definition}

\theoremstyle{remark}

\renewcommand{\theequation}{\arabic{equation}}
\begin{document}

\title{A d-summable approach to Deng information dimension of complex networks}

\author{Aldo Ramirez-Arellano}
{\address{SEPI-UPIICSA. Instituto Polit\'ecnico Nacional. Mexico City. M\'exico, C.P. 08400.}
\email{aramirezar@ipn.mx}}

\author{Juan Bory--Reyes}
\address{Escuela Superior de Ingenieria Mec\'anica y El\'ectrica\\
Instituto Polit\'ecnico Nacional\\
Edif. 5, 3er piso, U.P. Adolfo L\'opez Mateos\\
07338, Mexico City, MEXICO}
\email{juanboryreyes@yahoo.com}

\begin{abstract}
Several new network information dimension definitions have been proposed in recent decades, expanding the scope of applicability of this seminal tool. This paper proposes a new definition based on Deng entropy and d-summability (a concept from geometric measure theory). We will prove to what extent the new formulation will be useful in the theoretical and applied points of view. 
\end{abstract}



\maketitle

\section{Introduction}
The geometric dimension is a basic but important topic in complex networks and has received considerable attention as a fundamental quantity to characterize the structure and different physical properties of complex networks \cite{Law, Bun}. Since the beginning, it was applied to describe dense sets, like the points on a curve, surface, or volume, which belong to the fundamental methods of fractal geometry. In the past years, the body of research has focused on the calculation of the fractal dimension for complex networks, such as the volume dimension \cite{Shan,Guo,Dai2,Li}, correlation dimension \cite{Lac}, information dimension \cite{Dai}, local \cite{WEN20205} and multi-local information dimension \cite{WEN202001}, and Boulingand-Minkowski \cite{DESA2022} dimension, to name a few. The approaches to calculating the fractal dimension based on the box-covering method are described in \cite{Cha2,Sch,Kim,Wei2}. 

Information dimension of complex networks, introduced by Renyi based on the probability method \cite{Renyi2, Zhu-Ji2011}, has encouraged the introduction of d-summable information dimension \cite{RAMIREZARELLANO2019} and those based on Tsallis entropy \cite{Qi2016,Qi}, such as box-covering Tsallis information dimension \cite{RAMIREZARELLANO2020} and Renyi entropy \cite{DUAN2019}. There are many other network information dimensions in the literature, for example, adopting a fractional measure of entropy \cite{Ramirez2020chaos, RAMIREZARELLANO2021}. 

Most of the abovementioned information dimensions are based on the measure of entropies, which represent one parametric family of functionals to describe the uncertainty or randomness of a given system. The parameter value has been computed basically on the betweenness centrality measure \cite{Qi2016} and inner and outer degree \cite{RAMIREZARELLANO2021}, as well as the fraction of the minimum number of boxes to cover the network \cite{RAMIREZARELLANO2020}. Conversely, to calculate the Renyi information dimension’s parameter value, the approach restricts it to a fixed range \cite{DUAN2019}. A remarkable property of Tsallis \cite{Qi2016,RAMIREZARELLANO2020}, Renyi \cite{DUAN2019}, fractional \cite{Ramirez2020chaos,RAMIREZARELLANO2021} and d-summable information dimensions \cite{RAMIREZARELLANO2019} is that all of them converge to the classical information dimension (based on Shannon entropy) when their parameters tend to one.

Deng entropy \cite{DENG2016} was introduced in Dempster–Shafer’s theory (DST for brevity) framework, has recently gained attention. Several new measures of uncertainty as been derived from it, such as fractional Deng entropy \cite{Kazemi2021}, fractional Tsallis-Deng entropy \cite{BALAKRISHNAN2022}, and belief entropy \cite{Dan2019}, information. The information volume of a basic probabilistic assignment was introduced in\cite{Deng2020A,GAO2021,Chenhui2022} as the counterpart of a probability distribution information volume. Furthermore, an information dimension of complex networks based on Deng entropy was introduced in \cite{LEI2022}. The reader is referred to \cite{Ros2020, Ros2018, WEN2021} for some survey on the subject. 

For various reasons, it is reasonable to expect that to measure the topological structure and fractality of complex networks; it should be attainable to assume some minimum regularity of the form relative to the box dimension and information dimension. That is the case of the d-summable dimension, which has found applications in areas such as quantification of lung illness severity \cite{OrtizVilchis2022} and differentiation of lung lesions \cite{ramirez2021role}.

This paper reformulate the Deng information dimension of complex networks and show that the reformulated functional fit best synthetic and real-world complex networks. The remainder of this paper is organized as follows: in the next section, we briefly present the basic facts about Deng entropy, and Deng information dimension. Afterward, we will reformulate the Deng information dimension of complex networks. Finally, the findings of the obtained results and conclusion are discussed.

\section{Preliminaries}

\subsection{Deng entropy}
In Dempster-Shafer’s evidence theory, the uncertainty simultaneously contains non-specificity and discord, which coexists in Basic Probability Assignment (BPA for short). Let $X$ be  a set of $N$ mutually exclusive  and collective exhaustive events, indicated by  $X=\{\theta_1,\theta_2,..., \theta_i,...,\theta_N\}$, where set X is named a frame of discernment. The power set of 
$X$ is $$2^X=\{\emptyset,...,\{\theta_1\},..., \{\theta_N\}, \{\theta_1,\theta_2\},...,\{\theta_1,\theta_2,...,\theta_i\},...,X    \}$$

Given a frame of discernment $X$, the mass function, denoted by $m$, is a mapping from $2^X$ to [0,1]. It means that the basic probabilities of the elements in the power set $2^X$ are mapped to [0,1]. The mass function is also called BPA by imposing the condition:
\begin{equation}
\label{mfcondition}
m(\emptyset)=0, \sum_{A\subseteq 2^X} m(A)=1. 
\end{equation}

Let $m_1$ and $m_2$ be two BPA; the Dempster's combination rule is used just put the pieces together \cite{Dempster1967,SHAFER1976}:
\begin{equation}
\label{mcombination}
m(A)=\frac{1}{1-K}\sum_{B \cap C=A} m_1(B)m_2(C),
\end{equation}
where $K$ is called the conflict coefficient of the two BPAs, and it is given as:

\begin{equation}
\label{Kcombination}
K=\sum_{B \cap C=\emptyset} m_1(B)m_2(C).  
\end{equation}
Note that Eq.(\ref{mcombination}) is only applicable if $K<1$. 

Deng entropy \cite{DENG2016} as a uncertainty measure of BPA is defined as follow: 
\begin{equation}
\label{Dengentropy}
I_D=-\sum_{A \subseteq X} m(A) \log_2 \frac{m(A)}{2^{|A|}-1},  
\end{equation}
where $m$ is a mass function defined on the frame of discernment $X$, and $A$ is a focal element of $m$, meanwhile $|A|$ denotes $A's$ cardinality.  

Deng entropy is a generalization of celebrated Shannon entropy, and it is recovered when the BPA is a probability distribution. The difference between Deng and Shannon’s entropy is that belief of each focal element is divided by the term $2^{|A|-1}$, which represents the potential number of states in $A$. From Eq.(\ref{Dengentropy}), the term $\sum_{A \subseteq X} m(A) \log_2 (2^{|A|}-1)$ is the measure of the total non-specificity in the mass function $m$, and $-\sum_{A \subseteq X} m(A) \log_2 m(A)$ is the measure of discord of the mass function among various focal elements.

\subsection{Deng entropy information dimension}
\begin{figure}[hbt]
\centering
   \includegraphics[scale=0.3]{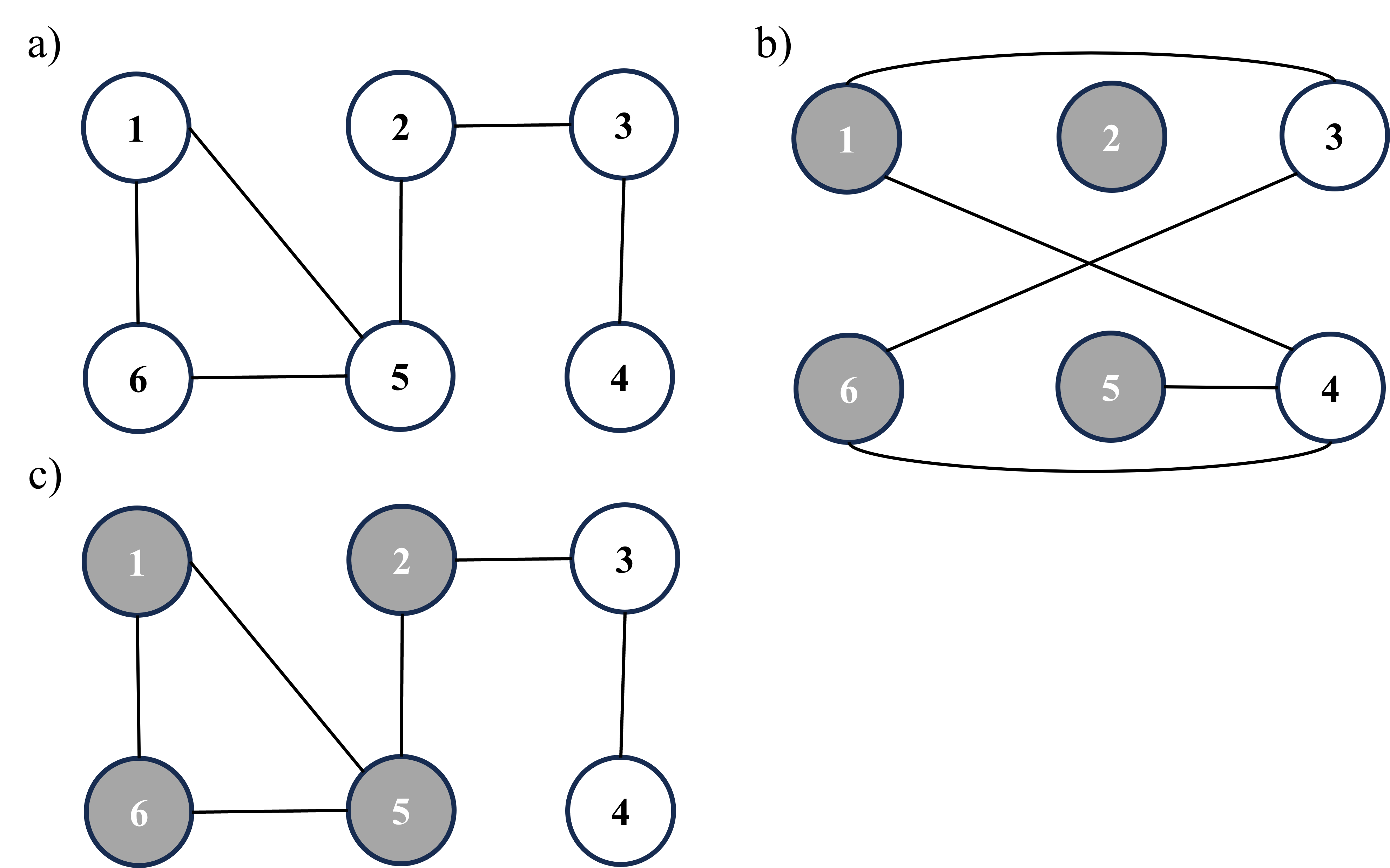} 
   \caption{Box-covering of a network for $\varepsilon=3$. Nodes of the same colour belong to the same box, $N_b=2$.}
    \label{fig1}
\end{figure}

The general information dimension can be defined as follows:
\begin{equation}\label{eginformationdimesnion}
d_{I}=-\lim_{\varepsilon \to 0}\frac{I}{\log \varepsilon} ,
\end{equation}
where $I$ is a generic measure of entropy and $\varepsilon$ is the diameter of the boxes to cover the network. The information dimension \cite{Dai} is obtained from Eq.(\ref{eginformationdimesnion}) by $I=-\sum p_i \log p_i$ that is Shannon entropy, while adopting $I_D$ of Eq.(\ref{Dengentropy}) yields
the Deng entropy information dimension \cite{LEI2022}:
\begin{equation}\label{eDenginfodimension}
d_{D}=\lim_{\varepsilon \to 0}\frac{\displaystyle\sum^{N_b}_{\substack{A_i \subseteq X \\ i =1} } m(A_i) \log_2 \frac{m(A_i)}{2^{|A_i|}-1}}{\log \varepsilon}. 
\end{equation}
Since network nodes are mutually exclusive, exhaustive collective events; the nodes are the elements of the frame of discernment $X$.  The focal element $A_i$ is set of nodes in the $i^{th}$ box of $N_b$ boxes discovered by the box-covering algorithm \cite{Cha}, and $|A_i|$ is its size, in our case, the number of nodes that belong to the box $A_i$. Note that $N_b$ depends on $\varepsilon$; hence, the focal elements (boxes) $A_i$ are generally constituted by different nodes for each $\varepsilon$. For example, when $\varepsilon=1$ each $A_i$ contains only one node; however, for $\varepsilon=3$, $A_1=\{1,2,5,6\}$ and $A_2=\{3,4\}$, see Figure \ref{fig1}c). 

The nodes connected by the shortest path in the network of the panel a) of Figure \ref{fig1} within a distance $\varepsilon=3$, the size of the boxes, should be connected in the graph in panel b). For example, node six is connected to three and four by a path of length three and four, respectively; hence they are connected in the network on panel b). Then, the colours of each node are assigned following the rule that two nodes connected in the network of panel b) not be in the same colour. Also, the number of colours of the nodes should be the minimum. Finally, the colours assigned in panel b) are mapped to the original network, see panel c). The boxes in the same colour belong to the same box, and the number of colours gives the number of boxes.

By (\ref{eDenginfodimension}), we can assert that
\begin{equation}\label{fDenginfo}
I_D(\varepsilon) \sim -d_{D} \log \varepsilon + \beta,
\end{equation} 
The mass function  or BPA  is given by:     
\begin{equation}\label{massfucntion}
m(A_i)=\frac{|A_i|}{N}, 
\end{equation}
where $|A_i|$ is the number of nodes of $A_i$ and $N$ is the number of network nodes. Since the box-covering algorithm \cite{Cha} does not allow overlapping boxes, each $A_i$ is a mutually exclusive subset of the nodes in the network; thus, $ \sum_{A_i \subseteq 2^{|X|}}  m(A_i)=1$.

\section{d-summable Deng information dimension}
A bounded and closed set is said to be d-summable if and only if the improper integral 
\begin{equation}\label{inted-summable}
\int_0^1 N_b(\varepsilon) \, \varepsilon^{d_d-1} \, d\varepsilon,
\end{equation}
converges. $N_b(\varepsilon)$ is the number of boxes of diameter $\varepsilon$ covering a compact set, and $d_d$ is the fractal dimension, named d-summable fractal dimension. The reader is referred to \cite{BV} for a detailed discussion of d-summable sets.

The d-summable dimension \cite{RAMIREZARELLANO2019} is the exponent $d_d$, such as the fractal relation
\begin{equation}\label{edfd1}
\frac{N_b(\varepsilon)}{\varepsilon^{1-\nu}} \sim C \varepsilon^{-d_d},
\end{equation} 
where $d_d$ is the d-summable dimension, and $ 0 < \nu < 1$ is a sufficient condition to warranty dsummability. Eq.(\ref{edfd1}) can be rewritten as:
\begin{equation}\label{edfd2}
d_d= -\lim_{\varepsilon \to 0} \frac {\log \frac{N_b(\varepsilon)}{\varepsilon^{1-\nu}}}{\log \varepsilon}.
\end{equation} 

Combining Eq.(\ref{eginformationdimesnion}), Eq.(\ref{edfd1}) with Eq.(\ref{edfd2}), we get a general d-summable information dimension:
\begin{equation}\label{gd-summabledinformation}
d_{dI} =-\lim_{\varepsilon \to 0}\frac{I}{\varepsilon^{1-\nu}\log{\varepsilon}},
\end{equation}
where $\varepsilon$ is the size of the boxes, and $I$ is a generic measure of entropy. If $I$ is replaced by Shannon entropy, we obtain what is known as the d-summable information dimension \cite{RAMIREZARELLANO2019}. It is clear that $d_{dI}$ reduces to $d_{I}$ when $\nu \rightarrow 1$.

We are ready to introduce the d-summable Deng information dimension of complex networks at this point. Replacing $I$ in Eq.(\ref{gd-summabledinformation}) for $I_D$ from Eq.(\ref{Dengentropy}), it makes the definition of the d-summable Deng information dimension allowable:
\begin{equation}\label{dengd-summabledinformation}
d_{dD} =\lim_{\varepsilon \to 0}\frac{\displaystyle\sum^{N_b}_{\substack{A_i \subseteq X \\ i =1}} m(A_i) \log_2 \frac{m(A_i)}{2^{|A_i|}-1}}{\varepsilon^{1-\nu}\log{\varepsilon}},
\end{equation}
where $\varepsilon$ is the diameter of the boxes to cover the network, $X$ is the frame of discernment formed by the network nodes. $A_i$ is the set of nodes in $i^{th}$ box discovered by the box-covering algorithm \cite{Cha}, and $|A_i|$ is the number of nodes that belong to the box $A_i$. 

The d-summable Deng information dimension naturally leads to the Deng information dimension when $\nu \rightarrow 1$. However, the practicability of Eq.(\ref{dengd-summabledinformation}) is doubtful since a indeterminnacy is obtained when  $\varepsilon=1$. This indeterminacy can be solved following \cite{Ros} by restricting the box size to $[2,\Delta-1]$, which $\Delta$ is the maximum size needed to cover the set with one box fully and $m(A_i)$ is the mass function of Eq.(\ref{massfucntion}).
 
Similarly, for some constant $\beta$, we see that 
\begin{equation}\label{fDengdsummableinfodimension}
I_{dD}(\varepsilon) \sim (-d_{dD} \log \varepsilon + \beta) \varepsilon^{1-\nu},
\end{equation} 
where $d_{dD}$ is the d-summable information dimension, and $\varepsilon$ is the size of the boxes needed to cover the set. 

In this context, the BPA measures each box's volume. Thus, the Deng entropy quantifies the information volume and uncertainty. In the limit case $\varepsilon=1$, the non-specificity $\sum_{A \subseteq X} m(A) \log_2 (2^{|A|}-1)=0$ and the discord  $-\sum_{A \subseteq X} m(A) \log_2 m(A)=-\log_2(\frac{1}{N})$ since $m(A)=\frac{1}{N}$, $|A|=1$; $\forall A \subseteq X$.
On the other hand, for $\varepsilon=\Delta+1$, the non-specificity is $\log_2(2^N-1)$ and the discord is zero since $m(A)=1$, $|A|=N$; $\forall A \subseteq X$. The non-specificity in the last case is large since it quantifies the number of partitions that can be built with $1,2,...N$ nodes. For example, assume a network with two nodes, three different partitions can be constructed: the node $n_1$ into partition $P_1$, the node  $n_2$ into box $P_2$;  $n_1$ into box  $P_2$ and $P_2$ into box $P_1$; finally the two nodes into one partition leading a non-specificity $\log_2(2^2-1)=\log(3)$ for $\varepsilon=\Delta+1$. For $\varepsilon=1$, each box contains one node; hence there is no other way to build a different box, leading to zero non-specificity. Additionally, the discord quantifies the information volume similar to the information dimension does.

The volume and the uncertainty of complex networks measured by Deng information dimension are reasonable to expect that it should be attainable to assume some minimum regularity expressed by Eq.(\ref{fDenginfo}). However, this regularity could not be presented in all complex networks; hence, the d-summable Deng information dimension measures the volume information dimension's non-regularity ($\nu$) as d-summability does on sets.

\section{Methodology}
The real-world networks used in the experimentation were bio-CE-PG (BCEPG), bio-grid-plant (BGP), bio-grig-worm (BGW), C. elegant neural network (CEN), dolphins social network (DS), Zachary’s karate club network (ZKC), power grid network (PG), protein interaction network (YEAST), USA airport network (USAA), E.coli cellular network (ECC), jazz-musician (JM), topology of a 1998-communications, network (TC), Lada Adamic’s social network (LAS), ca-netscience (CNC), email (EM) \cite{RAMIREZARELLANO2020}, see Table \ref{table1}. Also, small-world (SW) \cite{Watt} and scale-free networks based on Barabasi-Albert model (BA) were constructed, varying the nodes, see Table \ref{tablesin1}.  
 
For large networks with about 1500 nodes, the computation of $2^{|A|}-1$ is difficult since the precision of the CPU is not enough to get the exact number. A previous approach to solve this issue is to set the entropy to 0 when $|A|$ is too large \cite{LEI2022}. However, it disturbs the final entropy value. Hence, we propose to calculate both Deng and d-summable Deng information dimensions using $2^{|A|}$. This substitution simplifies Eq.(\ref{eDenginfodimension}) and approximates the entropy to the exact value. 

\begin{figure}[hbt]
\centering
   \includegraphics[scale=.2]{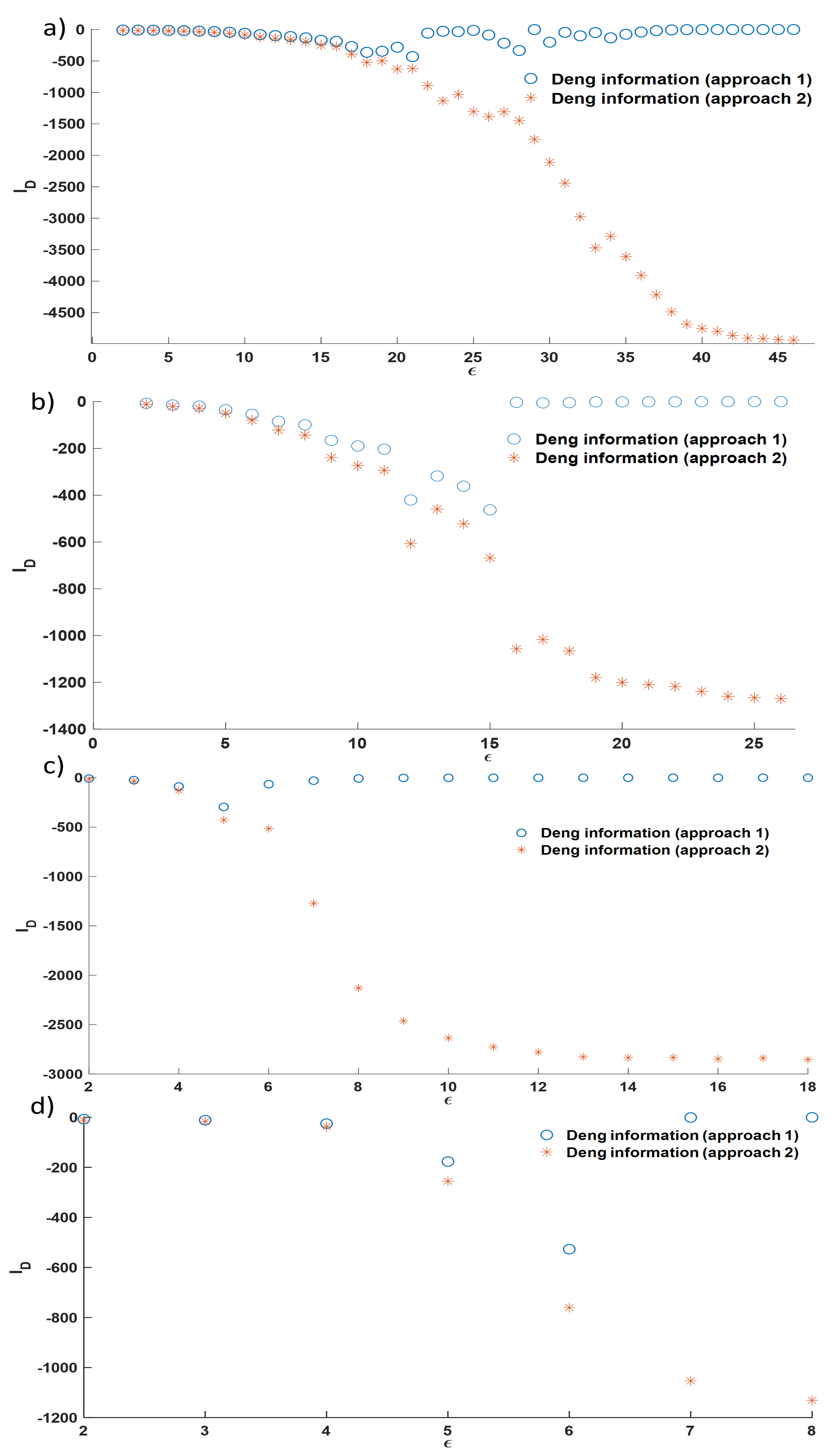} 
   \caption{ The proposed approach (star) and the previous one (circle) \cite{LEI2022} for the computation of $d_{D}$ for large $|A|$ on the a)PG, b)BGP, c)ECC and d)EM network.}
    \label{fig2}
\end{figure}

Figure \ref{fig2} shows the Deng information dimension computed using our approach (star) and the previous one (circle) \cite{LEI2022} on the a)PG, b)BGP c)ECC and d)EM network. The values of the Deng information dimension obtained by the previous approach and those obtained by the introduced here does not differ for low values of $\varepsilon$, but a high difference is found for large ones when the number of nodes in each box increase as much as $\varepsilon$. Our approximation was only used in networks where the exact value of Eq.(\ref{eDenginfodimension}) or Eq.(\ref{dengd-summabledinformation}) could not be obtained, such as the cases of BCEPG, BGP, BGW, ECC, EM, PG, and YEAST. 

The Deng information dimension ($d_D$) and d-summable Deng information dimension ($d_{dD}$),  as well as $\nu$, were calculated using a non-linear regression \cite{Seb} in MATLAB. Non-linear regression provides statistical evidence of the fit and complexity (number of parameters) of the models of Deng information Eq.(\ref{fDenginfo}) and d-summable Deng information Eq.(\ref{fDengdsummableinfodimension}) by the Akaike Information Criterion (AIC) \cite{Aka}. 

To provide a mean of model selection, the AIC of Eq.(\ref{fDenginfo}) and Eq.(\ref{fDengdsummableinfodimension}) are obtained; next, the min value is chosen ($AIC_{min}$). Finally, $\Delta AIC$ is computed by $AIC_{i}-AIC_{min}$, where $i$ is the $AIC$ of Eq.(\ref{fDenginfo}) and Eq.(\ref{fDengdsummableinfodimension}). The AIC's rule of thumb is that two models fit the data equally to each other if $\Delta AIC<2$ \cite{burn,burn2}. Otherwise, there is sufficient statistical evidence to select one model over the other.

\section{Results}

\subsection{Real-world network}

The features of the several real-world networks, Deng information dimension Eq.(\ref{eDenginfodimension}), and d-summable Deng information dimension Eq.(\ref{dengd-summabledinformation}) were computed, see Table \ref{table1}. The variation in the number of nodes and edges of real-world networks is sufficient representative.

\begin{table}[htb]

\caption{The nodes, edges,  Deng information dimension ($d_{D}$)  and d-summable Deng information dimension ($d_{dD}$, $\nu$)  of several real-world networks.}
\label{table1}
\centering
\begin{tabular}{|l|l|l|l|l|l|}
\hline

Network &Nodes & Edges &$d_D$	&$d_{dD}$&	$\nu$	
 \\ \hline 

BCEPG&1871&47754&1066.831&905.215&.919\\ \hline 

BGP&1745&3098&449.552&65.518&.351\\ \hline 

BGW&16347&6762822&7928.348&9637.435&1.082\\ \hline 

CEN&306&2359&147.728&12.902&-.56\\ \hline 

ECC&2859&6890&1197.332&948.451&.915\\ \hline 

EM&1113&10902&622.047&57.333&-.295\\ \hline 

DS&62&159&14.06&1.374&.081\\ \hline 


JM&198&2742&66.951&38.091&.704\\ \hline 

ZKC&34&78&8.632&8.524&.993\\ \hline 

LAS&350&3492&133.217&40.245&.4\\ \hline 

CNC&379&914&99.445&27.187&.515\\ \hline 

PG&4941&6594&1330.266&16.289&-.341\\ \hline 

TC&174&557&63.781&1.849&-.91\\ \hline 

USAA&500&5960&184.765&41.842&.234\\ \hline 

YEAST&2361 &7182&1093.266&149.067&.053\\ \hline 

\end{tabular}

\end{table}

Table \ref{table2} shows the adjusted coefficient of determination ($R^2$ adj), AIC of Deng information model Eq.(\ref{fDenginfo}) and d-summable Deng information model Eq.(\ref{fDengdsummableinfodimension}). These measures are denoted by the subscript ${D}$ and ${dD}$ for the Deng information model, and d-summable Deng information model, respectively. Since the d-summable Deng information model depends on $\nu$, the value of $\nu$ is also presented in Table \ref{table2} which was obtained by a non-linear regression \cite{Seb}. Note that $0<\nu<1$ for mostly real-world networks except for BGW, CEN, EM, PG, and TC.

\begin{table}[htb]
\caption{The $R^2$ adj, AIC, of the models for  Deng information (${D}$)  and d-summable Deng information (${dD}$, $\nu$)  of several real-world networks.}
\label{table2}
\centering
\begin{tabular}{|l|l|l|l|l|l|l|}
\hline

Network  &$R^2_{D}$ adj&	$R^2_{dD}$ adj&	$AIC_{min}$	&$\Delta AIC_{D}$	&$\Delta AIC_{dD}$& $\nu$	
 \\ \hline 
BCEPG&.947&.936&94.041&0&1.81&.919\\ \hline 
\textbf{BGP}&.807&.935&316.499&26.223&0&.351\\ \hline 
BGW&.871&.857&205.126&0&1.79&1.082\\ \hline 
\textbf{CEN}&.799&.925&37.476&4.728&0&-.56\\ \hline 
ECC&.875&.869&255.29&0&1.554&.915\\ \hline 
\textbf{EM}&.731&.911&92.066&7.309&0&-.295\\ \hline 
\textbf{DS}&.872&.925&36.872&3.251&0&.081\\ \hline 
JM&.87&.816&43.319&0&1.684&.704\\ \hline 
ZKC&.998&.996&.226&0&1.99&.993\\ \hline 
LAS&.841&.87&72.369&.963&0&.4\\ \hline 
\textbf{CNC}&.874&.939&149.573&10.839&0&.515\\ \hline 
\textbf{PG}&.618&.969&653.065&112.029&0&-.341\\ \hline 
\textbf{TC}&.688&.936&49.167&9.243&0&-.91\\ \hline 
\textbf{USAA}&.832&.851&66.148&.467&0&.234\\ \hline 
YEAST&.762&.905&145.08&8.582&0&.053\\ \hline 

\end{tabular}

\end{table}

\begin{figure}[hbt]
   \includegraphics[scale=.21]{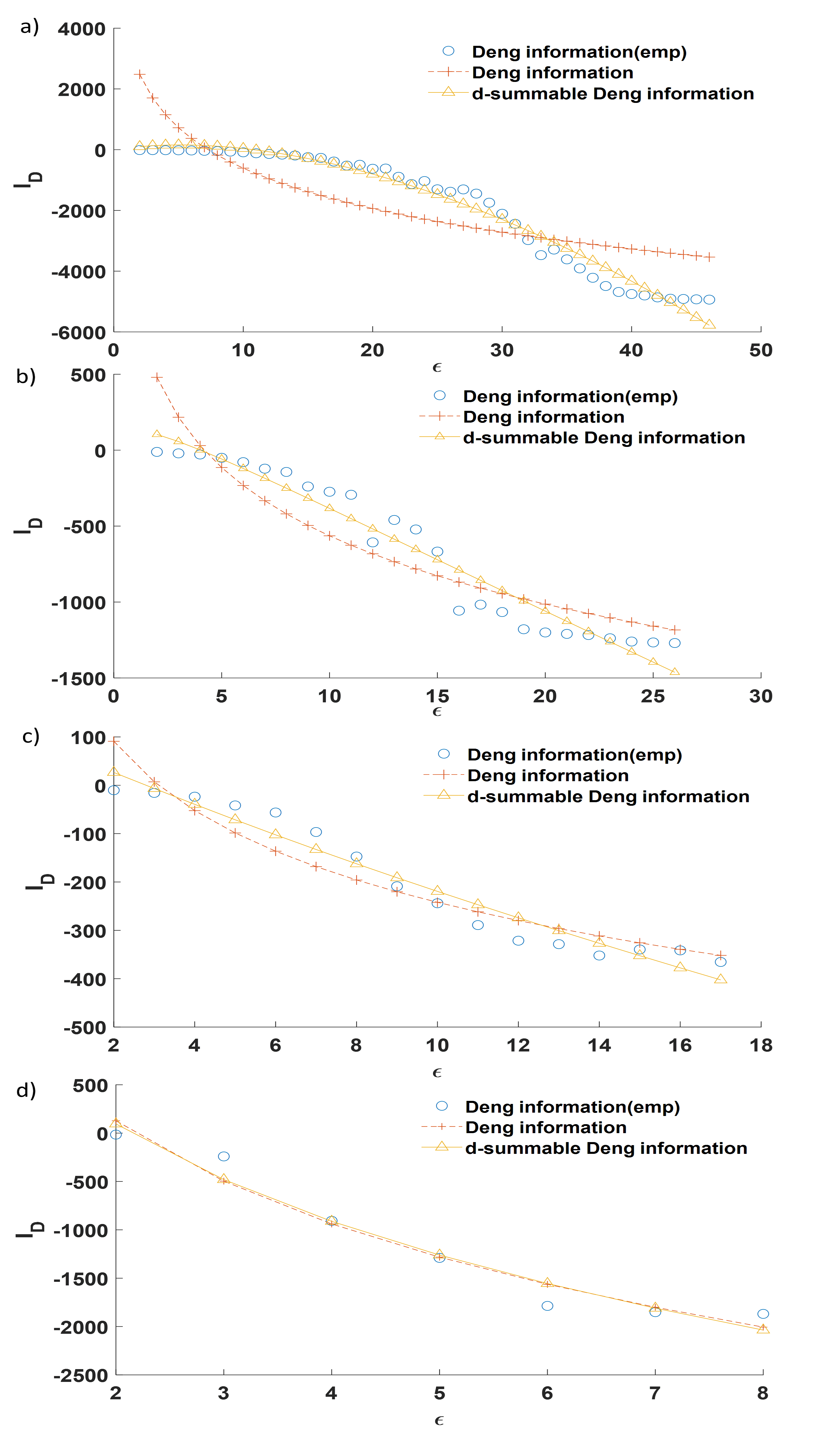} 
   \caption{ The fit of Deng information Eq.(\ref{fDenginfo}) and d-Summable Deng information Eq.(\ref{fDengdsummableinfodimension}) models on empirical Deng Information ( $\epsilon$ vs Deng entropy ) on a)PG ($\nu=-0.341$), b)BGP ($\nu=0.351$), c)CNC ($\nu=0.515$) and d)BCEPG ($\nu=0.919$). }
    \label{fig3}
\end{figure}

The d-summable Deng model best fits the empirical Deng entropy for eight out of 15 networks (in bold) since $\Delta AIC_{D} \geq 2 $, see Table \ref{table2}. The empirical Deng entropy is obtained from the computations on the real-world networks that are the points $\varepsilon$ vs Deng entropy, see Figure \ref{fig3}. For the remainder seven networks, there is no difference in the fit between the d-summable Deng information model and Deng information, see Table \ref{table2}, since $\Delta AIC_{dD} <2$ of BCEPG, BGW, ECC, JM, ZKC, LAS and USAA. Note that the higher value of $R^2$ adj coincides with the model that obtains $\Delta AIC=0$. However, for some networks, the $R^2$ adj values are closer; thus, drawing a convincing conclusion based only on it is rather difficult. Also, the $\nu$ values are related to the concavity and convexity of Eq.(\ref{fDengdsummableinfodimension}). The curve becomes convex as $\nu \rightarrow 1^{-}$, see panels a), b), c) and d) of Figure \ref{fig3}.

The d-summable Deng information and Deng information models fit equal for those networks; that $\nu$ value is closer to 1 as expected. Furthermore, for the networks that d-summable Deng information fitted better, $\nu$ is closer to 0 or even negative. The $d_{dD}$ value is less than $d_D$ for all real-world networks, even for those networks where both models fit equally. Hence, the d-summable Deng information dimension quantifies the topological features different to the Deng information dimension.

\subsection{Synthetic networks}

\begin{figure}[ht]
   \includegraphics[scale=.3]{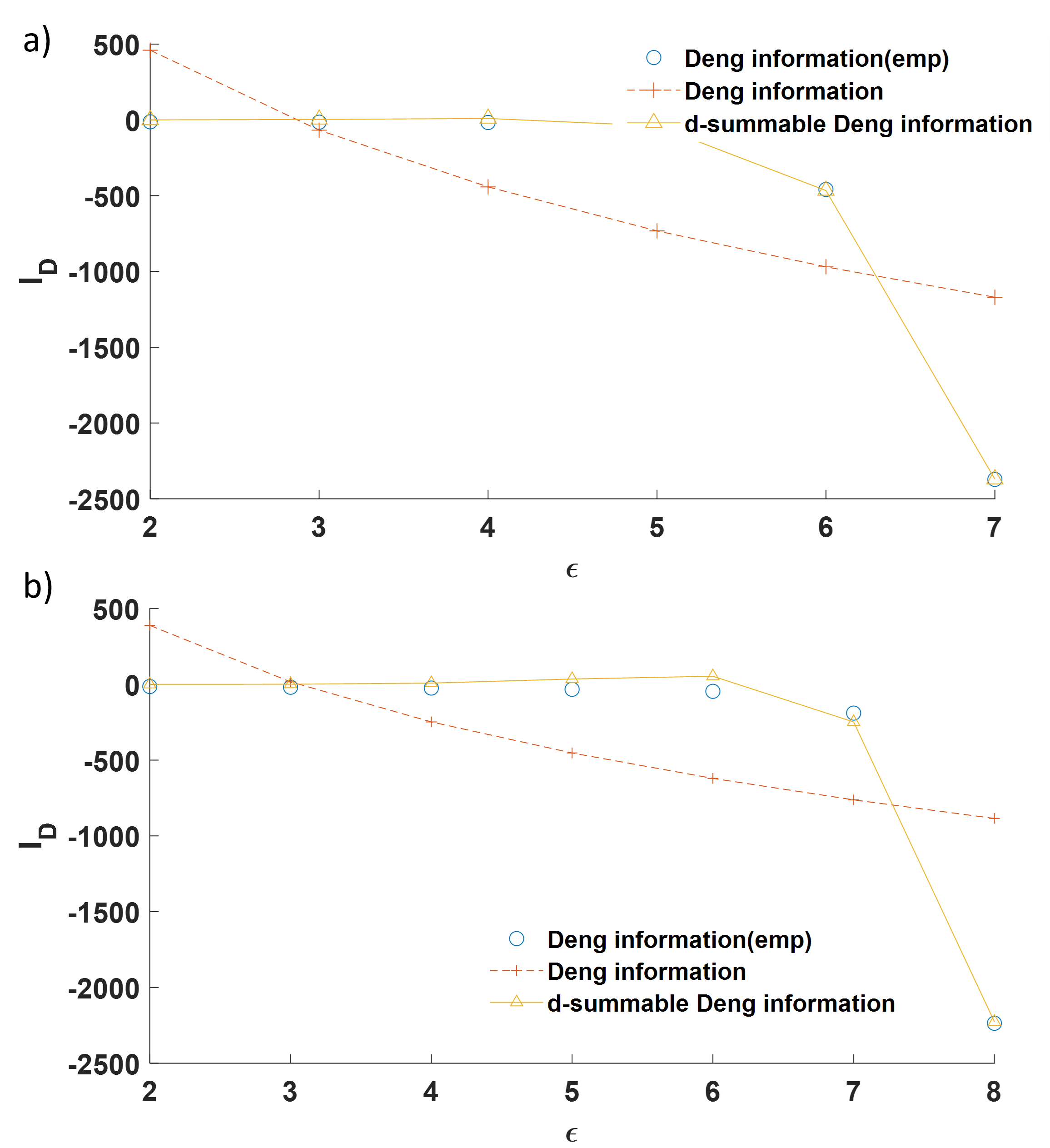} 
   \caption{ The fit of Deng information and d-Summable Deng information models on empirical Deng Information on a) BA-3000 ($\nu=-6.678$) and b)SW-3000 ($\nu=-8.167$). }
    \label{fig4}
\end{figure}

The SW \cite{Watt} and BA \cite{Bar} networks were constructed by varying the nodes, see Table \ref{tablesin1}. The Deng information dimension ($d_{D}$)  and d-summable Deng information dimension were computed on these networks. The $d_{dD}$ for synthetic networks is lower than $d_D$ and closer to 0, see Figure \ref{fig4}; this suggests that the d-summable Deng information dimension can differ between real-world and synthetic networks. Note that $d_{dD}$ of real-world networks are far from 0 and its value is not related to the network’s number of nodes and edges; see Tables \ref{table1} and \ref{tablesin1}. On the contrary, based on $d_D$, this differentiation is unclear. 
   
\begin{table}[ht]
\small
\caption{The nodes, edges,  Deng information dimension ($d_{D}$)  and d-summable Deng information dimension ($d_{dD}$, $\nu$) of several synthetic networks.}
\label{tablesin1}
\centering
\begin{tabular}{|l|l|l|l|l|l|l|}
\hline

Network &Nodes & Edges & $d_D$  & $d_{dD}$&$\nu$ \\ \hline 
BA-100&100&297&40.49&-9.00E-07&-4.087\\ \hline 
BA-200&200&598&51.609&1.57E-07&-5.889\\ \hline 
BA-500&500&1515&141.741&1.91E-04&-6.25\\ \hline 
BA-800&800&2422&266.108&1.801&-1.964\\ \hline 
BA-1000&1000&3014&315.472&.834&-2.507\\ \hline 
BA-1500&1500&7500&614.912&.139&-3.881\\ \hline 
BA-2000&2000&6058&732.029&.047&-4.575\\ \hline 
BA-3000&3000&9061&901.963&.001&-6.678\\ \hline 
SW-100&100&500&30.8&3.19E-06&-3.878\\ \hline 
SW-200&200&1000&58.077&4.09E-07&-4.078\\ \hline 
SW-500&500&2500&140.518&.001&-5.2\\ \hline 
SW-800&800&4000&183.591&9.44E-06&-8.267\\ \hline 
SW-1000&1000&5000&180.302&8.27E-06&-8.552\\ \hline 
SW-1500&1500&7500&528.252&.041&-4.207\\ \hline 
SW-2000&2000&10000&542.954&4.64E-05&-7.776\\ \hline 
SW-3000&3000&15000&637.021&3.78E-05&-8.167\\ \hline 

\end{tabular}

\end{table}

What is more, the d-summable Deng model better fits the empirical Deng Information of small-world and scale-free networks based on the AIC see Table \ref{tablesin2}. The $\Delta AIC_{dD}$ values are above 10, except for SW-100; thus, the evidence to prefer this model over the Deng information model is strong \cite{burn,burn2}. For synthetic networks, the $\nu$ values are lower than those of real-world networks.      

\begin{table}[ht]
\caption{The $R^2$ adj, AIC, of the models for  Deng information (${D}$)  and d-summable Deng information (${dD}$, $\nu$)  of several synthetic networks.}
\label{tablesin2}
\centering
\begin{tabular}{|l|l|l|l|l|l|l|}
\hline

Network  &$R^2_{D}$ adj&	$R^2_{dD}$ adj&	$AIC_{min}$	&$\Delta AIC_{D}$	&$\Delta AIC_{dD}$& $\nu$	
\\ \hline 
BA-100&.513&.985&24.464&11.933&0&-4.087\\ \hline 
BA-200&.395&.98&28.126&11.574&0&-5.889\\ \hline 
BA-500&.376&.997&34.094&27.199&0&-6.25\\ \hline 
BA-800&.507&.956&65.425&14.162&0&-1.964\\ \hline 
BA-1000&.467&.979&63.297&19.135&0&-2.507\\ \hline 
BA-1500&.381&.996&62.587&29.52&0&-3.881\\ \hline 
BA-2000&.345&.998&61.481&33.409&0&-4.575\\ \hline 
BA-3000&.269&.999&55.341&43.525&0&-6.678\\ \hline 
SW-100&.446&.857&28.798&6.193&0&-3.878\\ \hline 
SW-200&.494&.986&34.633&15.908&0&-4.078\\ \hline 
SW-500&.395&.994&46.947&27.174&0&-5.2\\ \hline 
SW-800&.249&.995&51.771&28.388&0&-8.267\\ \hline 
SW-1000&.216&.993&54.035&26.587&0&-8.552\\ \hline 
SW-1500&.335&.99&78.645&28.652&0&-4.207\\ \hline 
SW-2000&.209&.997&72.827&37.956&0&-7.776\\ \hline 
SW-3000&.159&.993&81.48&32.865&0&-8.167\\ \hline 

\end{tabular}

\end{table}

\section{Findings and conclusion} 
This article introduces a d-summable Deng information model (an extension of d-summable information dimension in DST). Our results show that the d-summable Deng information model fits better than the Deng information of the real-world and synthetic networks computed. This is confirmed by $AIC$ instead as a log-log plot that has been used extensively in previous research \cite{Guo,Dai2,Li,Lac,Dai,Cha2,Sch,Kim,Wei2}.

The d-summable Deng information model has a different functional form when $\nu \neq 1$; thus, $d_{dD}$ quantifies the information of the networks differently than $d_{D}$. On the other hand, for the networks where two models fit equally, the $d_{dD}$ is still lower than $d_{D}$. This empirical evidence supports the conjecture that $d_{D} > d_{dD}$. Furthermore, the $d_{dD}$ of real-world differs significantly from synthetic networks, where this value is closer to 0. In this case the d-summable term ($\varepsilon^{1-\nu} $) dominates the Eq.(\ref{fDengdsummableinfodimension}). This suggest that $d_{dD}$ is the slope of the plateau and $1-\nu$ is  the slope of the convex curve section, see Figure \ref{fig4}. The experimental calculation on synthetic and real-world networks leads to the belief that the above constitutes a successful identification criterion of these two kinds of networks when no additional information is available. Also, they support the conjuncture that Eq.(\ref{fDengdsummableinfodimension}) is concave when $\nu \leq 1/2$ (see Figures \ref{fig3} and \ref{fig4}), convex when $\nu>1/2$. Thus, when $\nu \rightarrow$ $1^{-}$ the concavity dilutes to give rise to convexity and the Deng information dimension is recovered.

\section*{Acknowledgements}
This research was partially supported by grant number 20230066 from SIP programs at Instituto Polit\'ecnico Nacional.

\bibliographystyle{unsrt}
\bibliography{references}

\end{document}